\documentclass[12pt,preprint]{aastex}
\shorttitle{Intermediate Mass Binary Pulsar Companions}
\shortauthors{Jacoby et al.}

\begin{document}
 
\title{Optical Detection of Two Intermediate Mass Binary Pulsar Companions}

\author{B. A. Jacoby\altaffilmark{1,2}, D. Chakrabarty\altaffilmark{3},
M. H. van Kerkwijk\altaffilmark{4}, S. R. Kulkarni\altaffilmark{1},
and D. L. Kaplan\altaffilmark{1,5}}

\altaffiltext{1}{Department of Astronomy, California Institute of
Technology, MS 105-24, Pasadena, CA 91125; srk@astro.caltech.edu.}

\altaffiltext{2}{present address: Naval Research Laboratory, Code 7213, 4555 Overlook Avenue, SW, Washington, DC, 20375; bryan.jacoby@nrl.navy.mil.}

\altaffiltext{3}{Department of Physics and Center for Space Research,
Massachusetts Institute of Technology, Cambridge, MA 02139;
deepto@space.mit.edu.}

\altaffiltext{4}{Department of Astronomy and Astrophysics, University
of Toronto, 60 St. George Street, Toronto, ON M5S 3H8, Canada;
mhvk@astro.utoronto.ca.}

\altaffiltext{5}{present address: Department of Physics and Center for Space Research, Massachusetts Institute of Technology, Cambridge, MA 02139;
dlk@space.mit.edu.}

\begin{abstract}

We report the detection of probable optical counterparts for two
Intermediate Mass Binary Pulsar (IMBP) systems, PSR~J1528$-$3146 and
PSR~J1757$-$5322.  Recent radio pulsar surveys have uncovered a
handful of these systems with putative massive white dwarf companions,
thought to have an evolutionary history different from that of the
more numerous class of Low Mass Binary Pulsars (LMBPs) with He white dwarf
companions.  The study of IMBP companions via optical observations
offers us several new diagnostics: the evolution of main sequence
stars near the white-dwarf-neutron star boundary, the physics of
white dwarfs close to the Chandrasekhar limit, and insights into the
recycling process by which old pulsars are spun up to high rotation
frequencies.  We were unsuccessful in our attempt to detect optical
counterparts of PSR~J1141$-$6545, PSR~J1157$-$5112,
PSR~J1435$-$6100, and PSR~J1454$-$5846.

\end{abstract}


\section{Introduction}\label{sec:intro}

The majority of recycled pulsars are in low mass binary pulsar (LMBP)
systems, consisting of a neutron star and a low-mass white dwarf.  The
LMBPs are widely considered to be descendants of the Low-Mass X-ray
Binaries (LMXBs).  The progenitors are thus a massive star primary
(which gives rise to the neutron star) and a low mass ($\lesssim\!1\,M_\odot$) 
secondary.  In contrast, double neutron star binaries,
exemplified by PSR~B1913+16, descend from binaries in which both the
primary and secondary are massive stars, each forming a neutron star.

Over the past few years, astronomers have come to appreciate the
existence of another class of binary pulsars, the so-called
intermediate mass binary pulsars (IMBPs) with massive C-O or O-Ne-Mg
white dwarf companions.  As suggested by their name, IMPBs are thought
to descend from binary star systems with a massive primary and a
secondary which is intermediate in mass.  First, the primary becomes a
neutron star through a supernova explosion.  Later, the secondary
evolves into a massive white dwarf, transferring matter to and
recycling the pulsar in the process \citep{vdh94, tvs00, tkr00}.  As
in LMBP systems, tidal damping circularizes the orbit because the
supernova occurs before the companion becomes a compact object.  16
candidate IMBP systems are currently known.

Not all pulsars with massive white dwarf companions share this
evolutionary path.  PSR~B2303+46 \citep{std85} and PSR~J1141$-$6545
\citep{klm+00a} have companions with masses similar to the IMBP systems;
however, these slowly-rotating pulsars appear to be unrecycled and
their orbits are eccentric.  In systems such as these it is thought
that neither the primary nor the secondary was initially massive enough
to form a neutron star.  As the primary evolved into a massive white
dwarf it transferred matter to the secondary, thereby making the
secondary massive enough to eventually become a neutron star.  Here
again the final outcome is a massive white dwarf and a neutron star,
but because the supernova occurs after the primary has become a white
dwarf the orbit remains eccentric and the neutron star is not recycled
\citep{ts00a}. Though only two such systems are known, they may exist
in numbers greater than neutron star binary systems \citep{py99}.  The
detection of the white dwarf companion via optical observations can
help clarify this interesting evolutionary path \citep{vk99}.

Apart from these tests of binary evolution, these systems may offer us
new insights into the physics of how neutron stars are spun up by
accretion.  It is clear that the mass transfer of the recycling
process results in a decreased magnetic field, as well as an increased
rotation rate for the neutron star.  The spin period at the end of the
spin-up phase, $P_0$, is a critical input to pulsar recycling models.
A comparison of the white dwarf age from cooling models with the
pulsar spin-down age (which assumes that $P_0$ is much smaller than
the current spin period) can, in principle, allow the determination of
$P_0$ \citep{ctk94}.

\section{Observations}

We have obtained optical observations of fields containing six IMBP
systems discovered in recent radio pulsar surveys with the Parkes
radio telescope (Fig. 1; Tab. 1; Camilo et al., 2001; Edwards \&
Bailes, 2001A; Kaspi et al., 2000). \nocite{clm+01, eb01, klm+00a} We observed
PSR~J1141$-$6545, PSR~J1157$-$5112, PSR~J1435$-$6100,
PSR~J1528$-$3146, PSR~J1454$-$5846, and PSR~J1757$-$5322 in $R$ band
on the nights of 6 -- 8 August 2002 with the the Magellan Instant
Camera (MagIC) on the 6.5\,m Baade telescope at Magellan Observatory.
Seeing was generally good, but some targets were observed at high
airmass, giving a broader point spread function.  Conditions were
photometric on 6 and 8 August, but there were clouds present on 7
August.  Each of our six targets was observed for two 10-minute
exposures on one of the photometric nights except for
PSR~J1528$-$3146.  These data were reduced following standard
practices (bias subtraction, flat fielding with dome flats),
photometrically calibrated with observations of the Stetson standard
star L112-805, and astrometrically calibrated using the USNO B-1.0
catalog.  The astrometric uncertainty in all observations presented
here is dominated by the tie between the USNO-B1.0 system and the
International Celestial Reference Frame ($\sim\!0\farcs2$ in each 
coordinate).

% Obs. dates were 7, 8, 9 August UT, 6/7, 7/8, 8/9 August local time.

On the night of 4 June 2003, we observed PSR~J1528$-$3146 once again
with MagIC.  Conditions were not photometric, but better than on our
previous attempt.  We obtained 2 exposures of 5 minutes each in $R$
and 2 exposures of 10 minutes each in $B$, reduced in the standard
manner as before.  A rough photometric calibration was obtained using
stars from the USNO B-1.0 catalog, which also provided the astrometric
calibration.  

% Obs. on 5 June UT, 4/5 June local.

Table \ref{tab:obs} gives the relevant parameters of the best imaging
observations in each band for each target. For each image, a model
point spread function was constructed based on several stars in the
field using the {\sc daophot} package in {\sc iraf}.  Limiting magnitudes were
determined by placing a number of artificial stars of a given
magnitude in the field and measuring their magnitudes with aperture
photometry.  This process was repeated to find the input artificial
star magnitude that resulted in a standard deviation of $\sim\!0.3$ in the
measured magnitude, corresponding to a 3\,$\sigma$ detection.

The second attempt at imaging the PSR~J1528$-$3146 field revealed a
faint object in the $R$ band image at $\alpha_{\rm J2000}=15^{\rm
h}28^{\rm m}34\fs955$, $\delta_{\rm
J2000}=-31\arcdeg46\arcmin06\farcs73$, and in the $B$ band image at
$\alpha_{\rm J2000}=15^{\rm h}28^{\rm m}34\fs945$, $\delta_{\rm
J2000}=-31\arcdeg46\arcmin06\farcs71$, consistent with the pulsar
timing position.  This potential counterpart is faint; we estimate $R
\sim 24.2$ and $B \sim 24.5$, but this photometry is somewhat
uncertain due to calibration with the USNO B-1.0 photographic
magnitudes.  This object is blue relative to most other stars in the
field.

Our observation of PSR~J1757$-$5322 showed a possible object at the
radio pulsar's timing position, but it was difficult to see in the
glare of a brighter star.  Subtraction of the brighter star from the
image using the {\sc daophot} {\sc substar} task reveals a
faint object with $R \sim 24.6$ at $\alpha_{\rm J2000}=17^{\rm
h}57^{\rm m}15\fs174$, $\delta_{\rm
J2000}=-53\arcdeg22\arcmin26\farcs17$, consistent with the pulsar
timing position.  We subsequently obtained a near-IR image of the
field with PANIC on the 6.5\,m Clay telescope at Magellan Observatory
on 18 April 2003, observing for a total of 72 minutes in $K_s$ band.
We subtracted dark frames, then produced a sky frame for subtraction
by taking a sliding box-car window of 4 exposures on either side of a
reference exposure.  We then added the exposures together, identified
all the stars, and produced masks for the stars that were used to
improve the sky frames in a second round of sky subtraction.
Astrometry was again provided by the USNO B-1.0 catalog, and
photometric calibration by comparison with several 2MASS stars in the
field.  There is no object present at the pulsar's position to the
detection limit of the image, $K_s = 20.8$.  The implied limit on the color
corresponds to a main sequence spectral type of $\sim\,$M4 or earlier,
and is thus consistent with a white dwarf.

Several of these fields are rather crowded; this was especially
problematic in the case of PSR~J1435$-$6100, whose position overlaps
with three blended objects in our image.  On the night of 6 June 2003,
we obtained a spectrum of the bright object near the pulsar position
with the LDSS2 on the Clay telescope, and determined that it is a
reddened F-type main sequence star and thus not associated with the
pulsar.  We used the {\sc daophot} {\sc allstar} task to subtract
stars near the positions of PSR~J1157$-$5112 and PSR~J1435$-$6100,
eliminating the possibility of fainter counterparts hidden by the
nearby brighter objects in these cases.

\section{Discussion and Conclusions}

We detected optical counterparts for two out of the six IMBP systems
we studied, PSR~J1528$-$3146 and PSR~J1757$-$5322.  From Table
\ref{tab:targets}, one sees that these are the two nearest targets.
Thus, it is quite possible that deeper observations would reveal
the counterparts in the remaining binaries as well.

In Figure \ref{fig:coolingcurve}, we show cooling curves for hydrogen
atmosphere white dwarfs with masses from 0.5\,$M_\odot$ to
1.2\,$M_\odot$, along with the observationally-inferred absolute $R$
magnitudes of massive white dwarf pulsar companions versus the
spin-down ages of their pulsars.  The absolute magnitudes have
large uncertainties which are difficult to quantify because the only
constraint on the pulsar distances is based on dispersion measure and
a model of the galactic electron distribution \citep{cl02}; however,
this exercise is still instructive.  We note that in all cases where
optical observations failed to detect an IMBP counterpart, the
predicted magnitude is fainter than the observation's detection
threshold.

As previously mentioned, it is thought that the companion stars in the
PSR~J1141$-$6545 and PSR~B2303+46 systems must have been fully evolved
by the time the pulsars formed.  Therefore, in these systems, the pulsar
age does not constrain the white dwarf age and the failure to detect
the PSR~J1141$-$6545 companion is not troubling.  The detected optical
counterpart of PSR~B2303+46 \citep{vk99} is significantly fainter than
predicted by the cooling model based on the pulsar's spin-down age.
In addition to the expectation that the white dwarf is older than the
pulsar, this object has the largest $z$-distance from the galactic
plane in this sample; it is above much of the ionized gas in the
galactic disk, so the dispersion measure-based distance estimate could
be significantly smaller than the true distance.  

In all of the other systems, the neutron star formed first and the
pulsar's spin-down age should, in principle, correspond to the time
since the end of the companion's evolution.  The other five detected
objects are all brighter than predicted by the cooling curves if they
are as old as their pulsars' characteristic ages.  Although there is a
large uncertainty associated with the absolute magnitude of each
object, as a group, they suggest that the standard spin-down model for
pulsars may in fact significantly overestimate the pulsar age in these
cases, possibly because $P_0$ was not much smaller than the
current spin period.

\acknowledgments

BAJ and SRK thank NSF and NASA for supporting their research.  MHvK
acknowledges support by the National Sciences and Engineering Research
Council of Canada.  DLK thanks the Fannie \& John Hertz Foundation for
its support. BAJ holds a National Research Council Research Associateship Award
at the Naval Research Laboratory.  Basic research in astronomy at NRL is 
supported by the Office of Naval Research.

%\bibliographystyle{apj} 
%\bibliography{modrefs,psrrefs}

\begin{deluxetable}{crcccrcrc}
\tabletypesize{\scriptsize}
\tablewidth{0pt}
\tablecaption{Parameters of six target massive white dwarf binary systems \label{tab:targets}}
%\tablehead{\colhead{Pulsar} & \colhead{$P$ (ms)} & \colhead{$B$ (G)} & \colhead{$\tau_c$ (Gyr)} & \colhead{$P_b$ (d)} & \colhead{$e$} & \colhead{$m_{\rm c\,min}$ ($M_\odot$)} & \colhead{$d$ (kpc)\tablenotemark{a}} & \colhead{Reference}}
\tablehead{\colhead{Pulsar} & \colhead{$P$} & \colhead{$B$} & \colhead{$\tau_c$} & \colhead{$P_b$} & \colhead{$e$} & \colhead{$m_{\rm c\,min}$} & \colhead{$d$\tablenotemark{a}} & \colhead{Reference} \\  & \colhead{(ms)} & \colhead{($10^9$\,G)} & \colhead{(Gyr)} & \colhead{(d)} &  & \colhead{($M_\odot$)} & \colhead{(kpc)} & }

\startdata
J1141$-$6545 & 393.9 & 1300 & 0.0014 & ~0.20 & $1.8\times10^{-1}$ & 0.97 & 2.5 & 1 \\
J1157$-$5112 & ~43.6 & 2.5 & 4.8 & ~3.51 & $4.0\times10^{-4}$ & 1.18 & 1.3 & 2 \\ 
J1435$-$6100 & ~~9.3 & 0.5 & 6 & ~1.35 & $1\times10^{-5}$ & 0.90 & 2.2 & 3 \\ 
J1454$-$5846 & ~45.2 & 6 & 0.9 & 12.42 & $1.9\times10^{-3}$ & 0.87 & 2.2 & 3 \\ 
J1528$-$3146 & ~60.8 & 3.9 & 3.9 & ~3.18 & $2.1\times10^{-4}$ & 0.94 & 0.80 & 4 \\ 
J1757$-$5322 & ~~8.9 & 0.49 & 5.3 & ~0.45 & $(4 \pm 4)\times10^{-6}$ & 0.55 & 0.96 & 2 \\ 
\enddata
\tablenotetext{a}{Distance estimated from dispersion measure using model of \cite{cl02}}
\tablerefs{(1) Kaspi et al., 2000; (2) Edwards \& Bailes, 2001b\nocite{eb01b}; (3) Camilo et al., 2001; (4) Jacoby et al., in prep.}
\end{deluxetable}

\begin{deluxetable}{ccccc}
\tabletypesize{\scriptsize}
\tablewidth{0pt}
\tablecaption{Observations of massive white dwarf binary systems\label{tab:obs}}
\tablehead{\colhead{Pulsar} & \colhead{Filter} & \colhead{seeing} & \colhead{Detection Limit} & \colhead{Potential Counterpart\tablenotemark{a}}\\
 &  & \colhead{(arcsec)} & \colhead{(magnitudes)} & \colhead{(magnitudes)}}
\startdata
J1141$-$6545 & $R$ & 1.1 & 23.4 & -- \\
J1157$-$5112 & $R$ & 1.2 & 23.7 & -- \\
J1435$-$6100 & $R$ & 1.0 & 23.1 & -- \\
J1454$-$5847 & $R$ & 0.8 & 24.9 & -- \\
J1528$-$3146 & $R$ & 0.7 & 24.4 & 24.2(4) \\
              & $B$ & 0.7 & 25.9 & 24.5(2) \\
J1757$-$5322 & $R$ & 0.6 & 24.8 & 24.6(2) \\
              & $K_s$ & 0.5 & 20.8 & -- \\
\enddata
\tablenotetext{a}{Figures in parenthesis are uncertainties in the last digit quoted.}
\end{deluxetable}

\begin{figure}
\includegraphics[angle=270,scale=0.27]{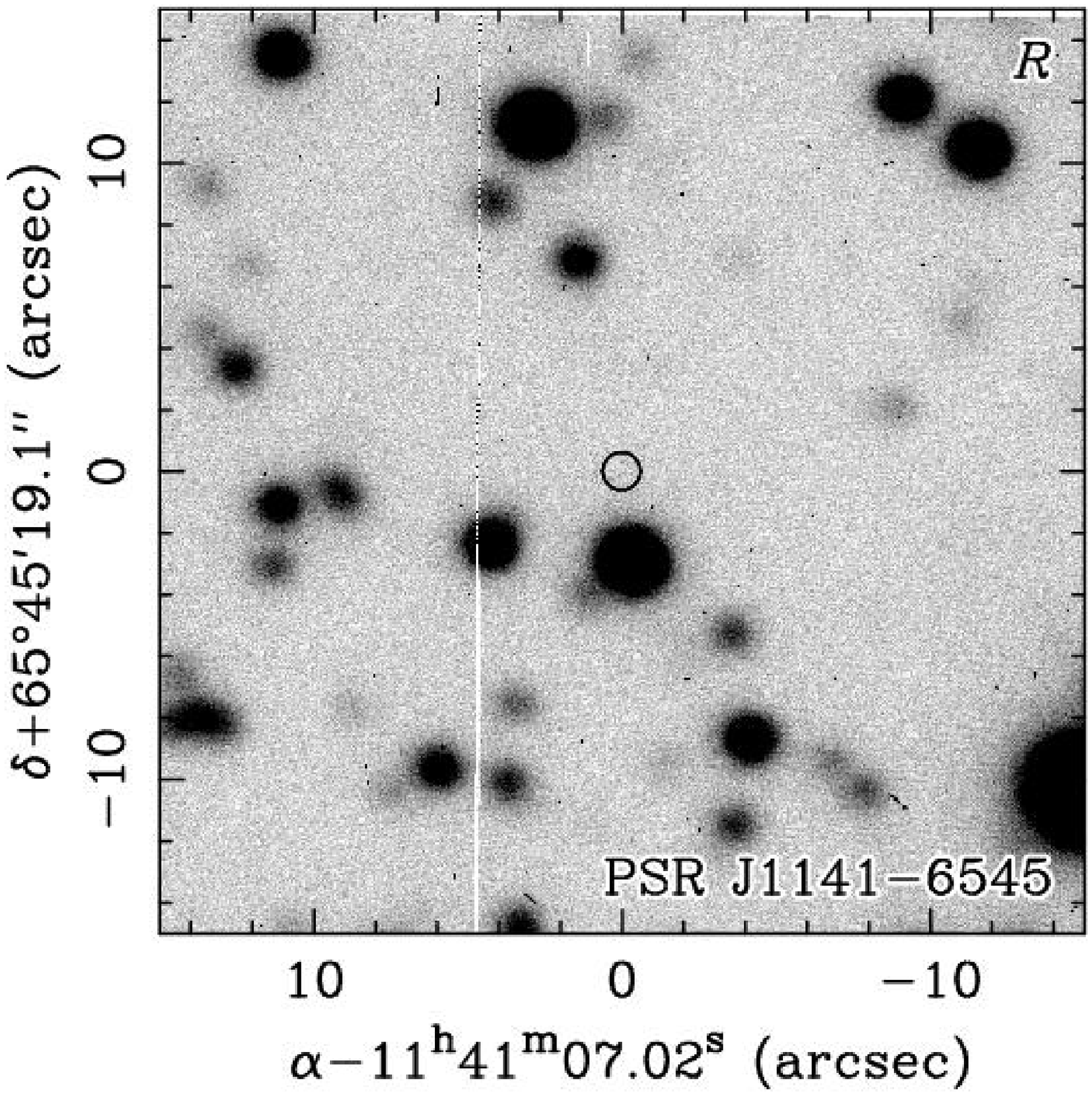} \hspace{1.5mm} \includegraphics[angle=270,scale=0.27]{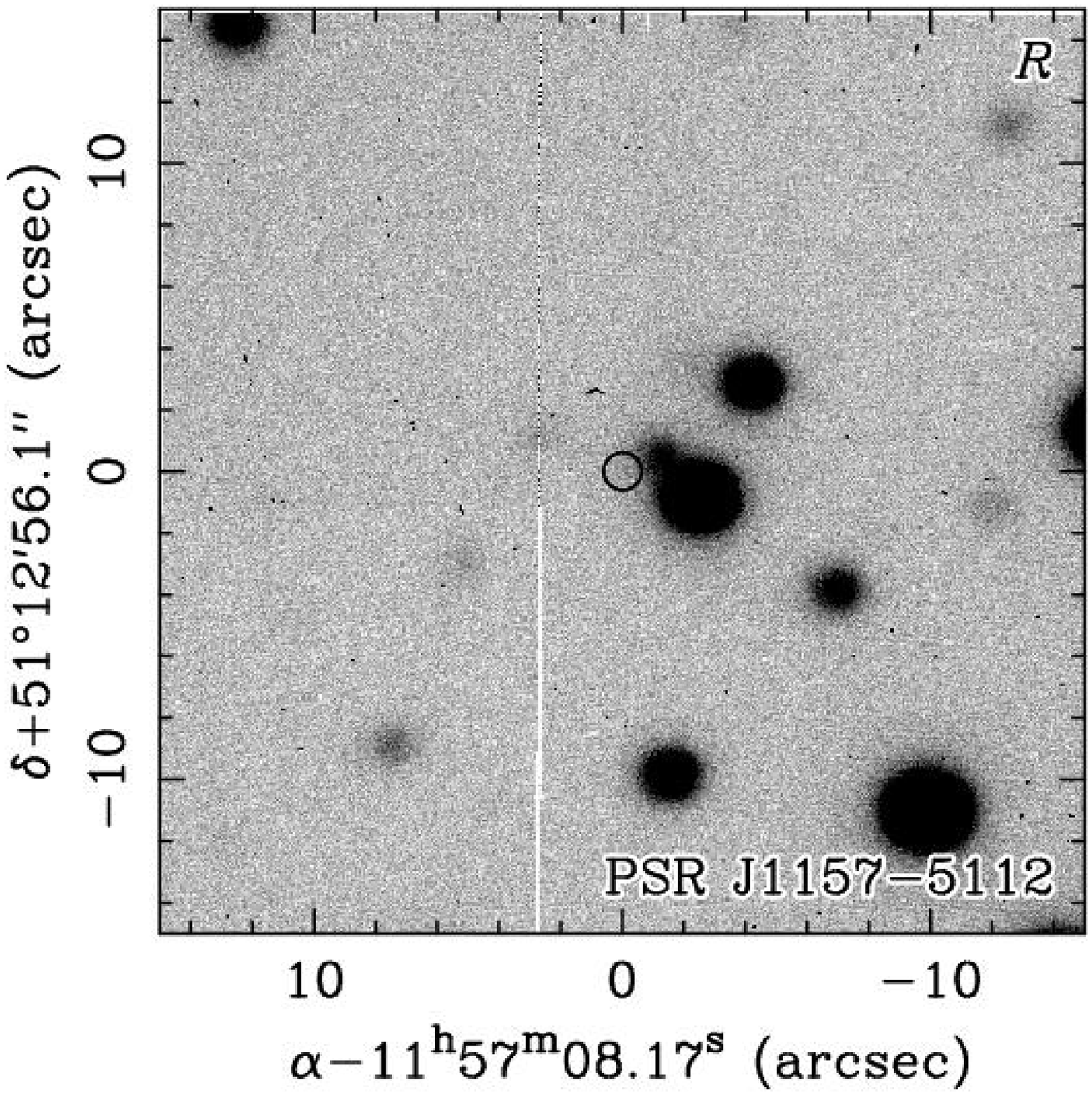} \hspace{1.5mm} \includegraphics[angle=270,scale=0.27]{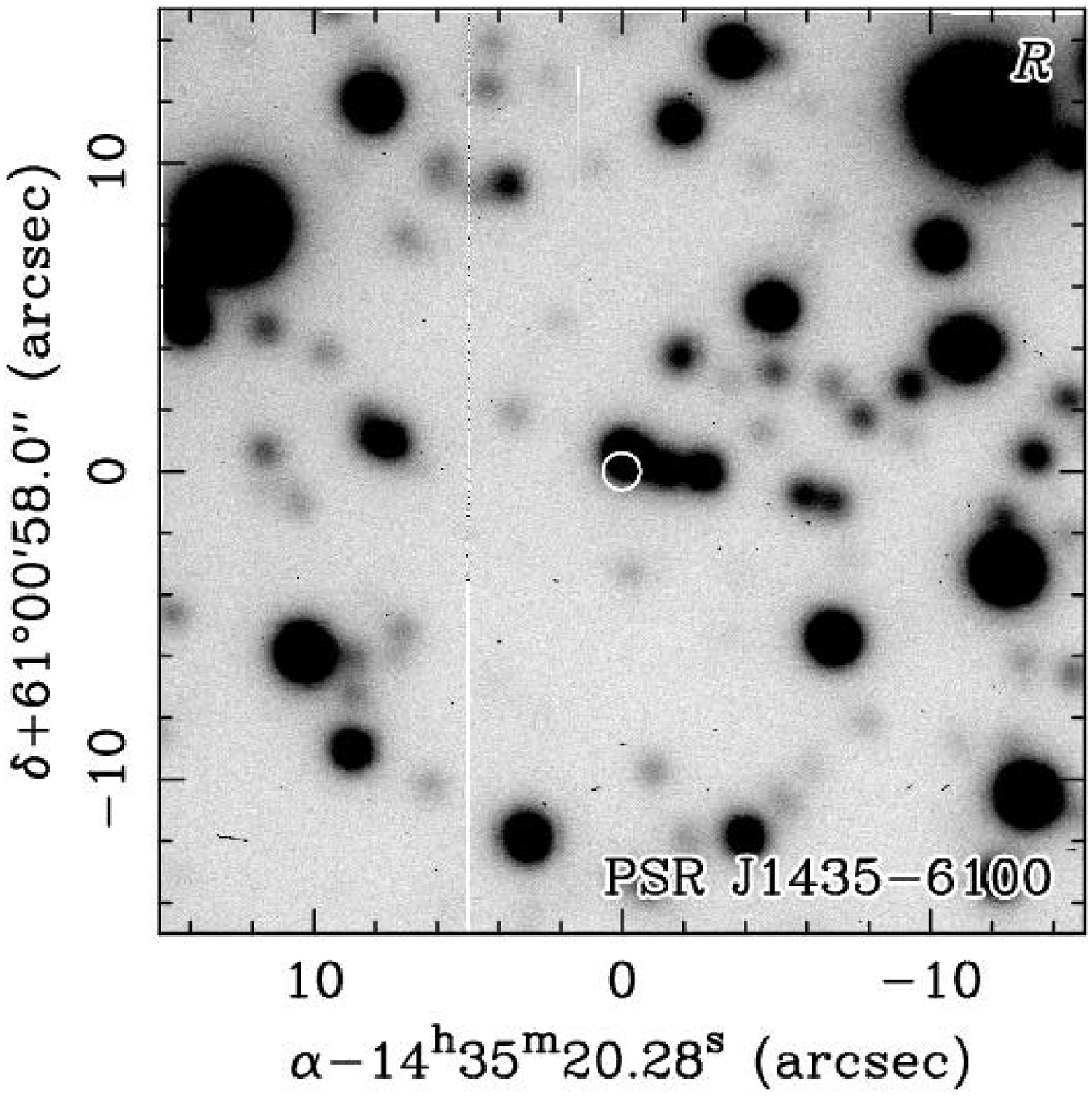} \\ \vspace{1.5mm} \\ \includegraphics[angle=270,scale=0.27]{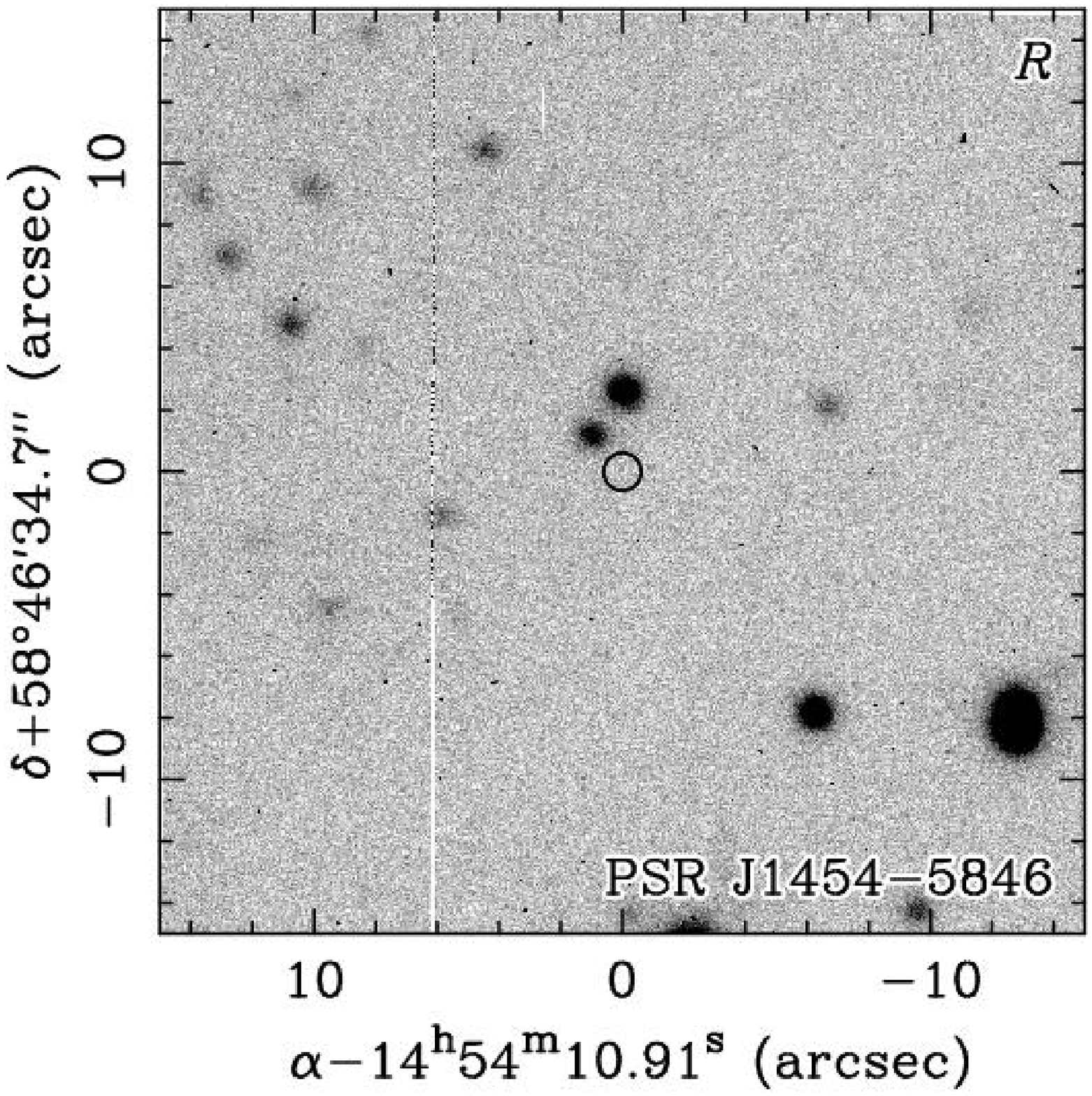} \hspace{1.5mm} \includegraphics[angle=270,scale=0.27]{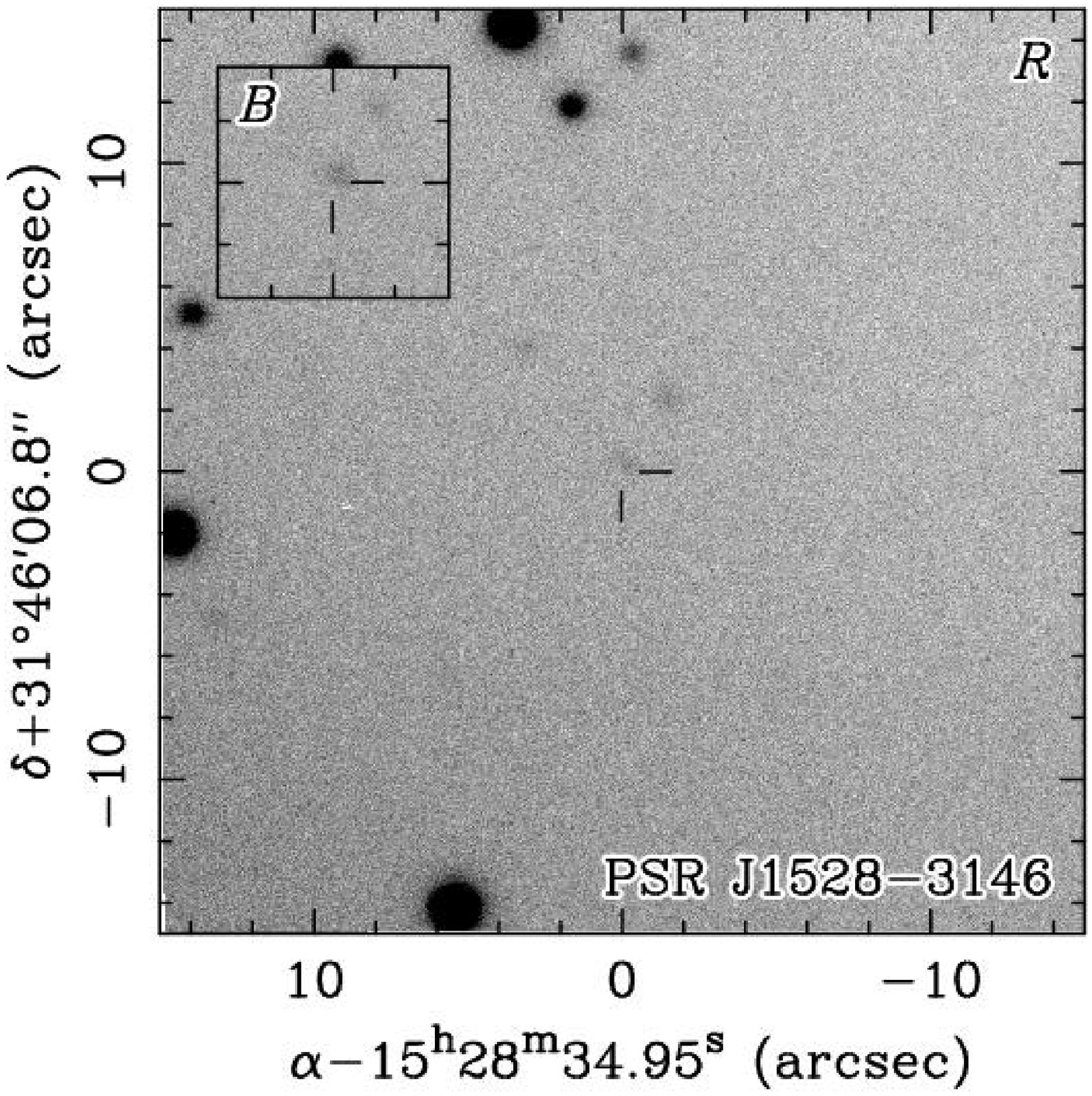} \hspace{1.5mm} \includegraphics[angle=270,scale=0.27]{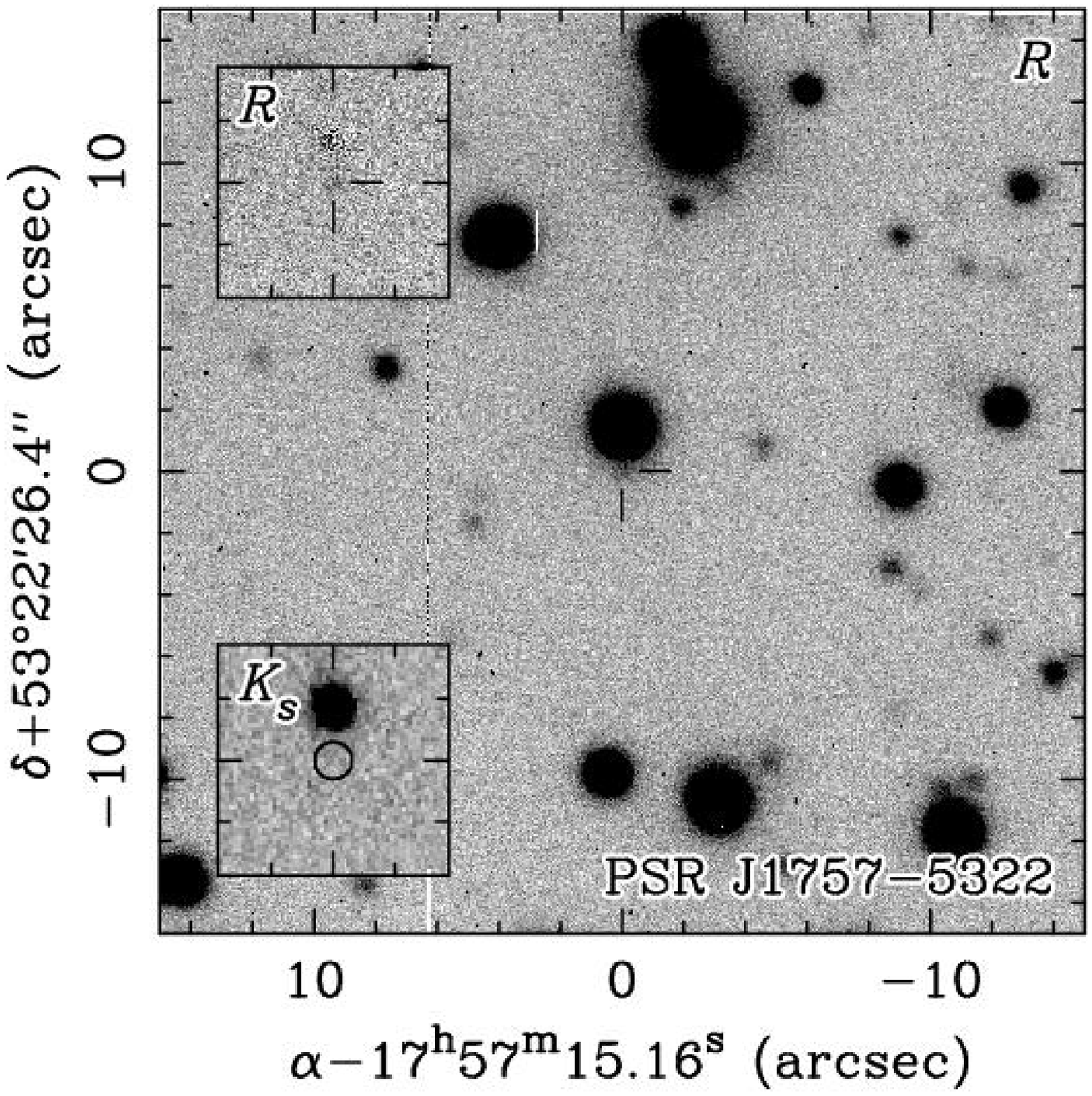}

\caption[]{Images of fields of PSR~J1141$-$6545, PSR~J1157$-$5112,
PSR~J1435$-$6100, PSR~J1528$-$3146, PSR~J1454$-$5846, and
PSR~J1757$-$5322.  Circles indicate the 3\,$\sigma$ uncertainty in the
pulsar position; tick marks show the pulsar position where a plausible
counterpart was detected.  Large images are in $R$ band.  For
PSR~J1528$-$3146, the inset shows the $B$ band image.  For
PSR~J1757$-$5322, the inset at upper left shows the $R$ band image
after the subtraction of bright star near pulsar position, and the
inset at lower left shows the $K_s$ image.  For PSR~J1141$-$6545, the
timing position from \citet{bok+03} was used; in all other cases
positions were taken from the references in Table \ref{tab:targets}.
\label{fig:images}}
\end{figure}

\begin{figure}
\hspace{3cm}\includegraphics[angle=0,scale=0.5]{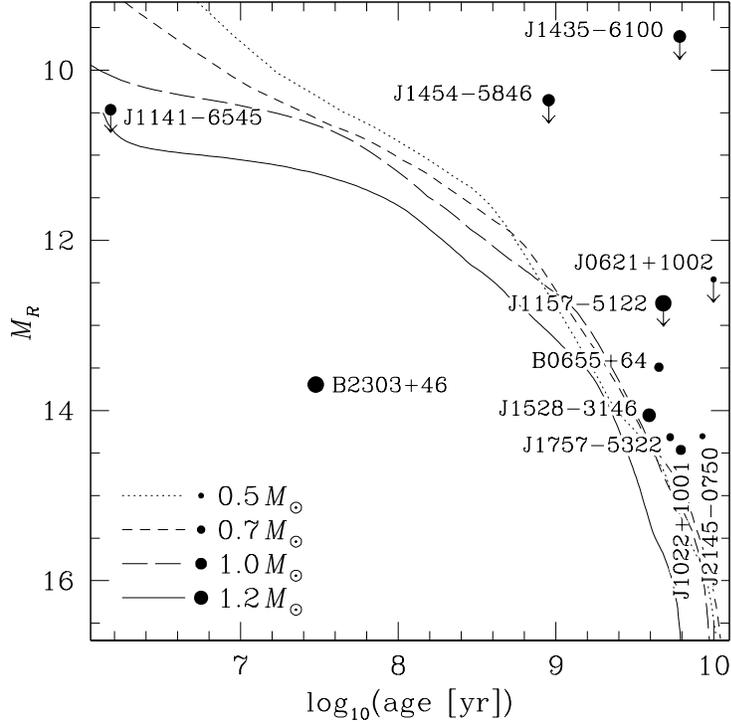}
\caption[]{White
dwarf cooling curves and observations of massive white dwarf pulsar
companions.  The curves show the absolute $R$ magnitude versus age for
massive white dwarfs with hydrogen atmospheres.  Points show the
observationally-derived $M_R$ versus pulsar characteristic age for
massive white dwarf pulsar companions with arrows indicating upper
limits from non-detections.  The diameter of each point is
proportional to the most likely mass of the white dwarf, assuming a
1.35\,$M_\odot$ pulsar and $60^\circ$ orbital inclination with these
exceptions: the companion of PSR~J1141$-$6545 has a mass of 
(0.99$\pm$0.02)\,$M_\odot$ \citep{bok+03}; the most likely masses of
the companions of PSR~B2303+46 and PSR~J1157$-$5112 are greater than the
Chandrasekhar mass, so we have assigned them diameters corresponding
to 1.4\,$M_\odot$.  The curves are based on the luminosity -- age
relation for 0.5, 0.7, and 1.0\,$M_\odot$ white dwarfs with hydrogen 
fractions of $10^{-4}$, and a 1.2\,$M_\odot$ white dwarf with a
hydrogen fraction of $10^{-6}$, all with zero metallicity for the envelope
\citep{ba99}.  To these cooling relations, we applied bolometric
corrections and colors as a function of temperature for a log(g) = 8
white dwarf with a hydrogen atmosphere \citep{bwb95}.  Apparent $R$
magnitudes were converted to absolute $R$ magnitudes using the
dispersion measure-distance model of \cite{cl02}, with an extinction
correction from \cite{nk80} for objects within $5^\circ$ of the
galactic plane, and from \cite{sfd98} for higher latitude pulsars.  In
the cases of PSR~J2145$-$0750 and PSR~B0655+64, $R$ was calculated
based on the measured $V$ and inferred temperature of \cite{lfc96},
using colors from \cite{bwb95}.  Photometry for PSR~J0621+1002 is from
\cite{kul86}; PSR~B2303+46 from \cite{vk99}; and PSR~J0621+1002 from
\cite{vbj+04}.}
\label{fig:coolingcurve}
\end{figure}

\end{document}